# II. The Standard Model in the Isotopic Foldy-Wouthuysen Representation without Higgs Bosons in the Fermion Sector. Spontaneous Breaking of Parity and "Dark Matter" Problems

V.P. Neznamov*

RFNC-VNIIEF, Sarov, Russia

Abstract

The Standard Model with massive fermions is formulated in the isotopic Foldy-Wouthuysen representation. $SU(2) \times U(1)$ - invariance of the theory in this representation is independent of whether fermions possess mass or not, and, consequently, it is not necessary to introduce interactions between Higgs bosons and fermions.

The study discusses a possible relation between spontaneous breaking of parity in the isotopic Foldy-Wouthuysen representation and the composition of elementary particles of "dark matter".

e-mail: neznamov@vniief.ru

# INTRODUCTION

This Paper II is a continuation of Paper I [1]. It deals with the construction of the Standard Model in the isotopic Foldy-Wouthuysen (IFW) representation. The second part of this study discusses the probable relation between spontaneous breaking of parity in the IFW representation and the problems and composition of "dark matter" particles.

As we know in the original Standard Model is considered massless fermions to ensure the $SU(2)$ invariance. Masses are imparted to fermions after the introduction of the mechanism of spontaneous breaking of symmetry, the resulting occurrence of Higgs bosons and postulation of their gauge-invariant interactions with fermions with Yukawa coupling [2].

Earlier, in Refs. [3], [4], the author considered the Standard Model in the modified Foldy-Wouthuysen representation. The $SU(2)$-invariant formulation of the theory in this representation was shown to be possible for originally massive fermions. In this case, it is not necessary to require that Higgs bosons should interact with fermions. Within this approach, fermion masses are introduced from outside. The theory has no vertices of Yukawa interactions between Higgs bosons and fermions; Higgs bosons are responsible only for the gauge invariance of the theory's boson sector and interact only with gauge bosons $W_\mu^\pm$, $Z_\mu$, gluons and photons.

Therefore, in the theory, there are no processes of Higgs boson decay to fermions $\left(H \to f \bar{f}\right)$, no quarkonium states $\Psi, \Upsilon, \theta$ including Higgs bosons, no interactions between Higgs bosons and gluons $(ggH)$ and photons $(\gamma\gamma H)$ via fermion loops, etc.

Because of the unitarity of the Foldy-Wouthuysen transformation, the remaining theoretical results of the Standard Model (including the theory's renormalization property obtained earlier in the Dirac representation are preserved.

The objective of this study is to construct the Standard Model in the isotopic Foldy-Wouthuysen representation introduced in previous Paper I [1]. Due to the chirally symmetric equations of fermion fields in the isotopic Foldy-Wouthuysen representation with massive fermions, the absence of necessity of introducing the interaction between fermions and Higgs bosons becomes particularly clear.

The second part of this study considers the possibility of spontaneous breaking of parity in the isotopic Foldy-Wouthuysen representation. As noted above, the Lagrangian and the Hamiltonian of the Standard Model in the isotopic Foldy-Wouthuysen representation are chirally symmetric, whether fermions have mass or not.

However, the fermion vacuum in the IFW representation is degenerate [1]. One vacuum state is a chirally symmetric state of the Dirac sea of negative-energy fermions.

Two other vacuum states are the states of the sea of negative-energy fermions with disturbed P - symmetry.

In the IFW representation, for fermions, this creates conditions for spontaneous breaking of P - symmetry. If this phenomenon is the case, there is the connection between broken P - symmetry with problems and composition of elementary particles of "dark matter".

The units and notations used in this paper are the same as in previous Paper I [1].

## 1. THE STANDARD MODEL IN THE ISOTOPIC FOLDY-WOUTHUYSEN REPRESENTATION

The Standard Model possesses $SU(3) \times SU(2) \times U(1)$ gauge symmetries [2]. In order to construct the Standard Model in the IFW representation, we first consider the electroweak Glashow-Weinberg-Salam (GWS) model [5], which is invariant with respect to the transformations of $SU(2) \times U(1)$. Quantum chromodynamics (QCD) possessing local $SU(3)$ symmetry can be easily introduced into the Standard Model in the isotopic Foldy-Wouthuysen representation after constructing the GWS model in the IFW representation. For simplicity, at the beginning we consider only the first generation of leptons $(v_e, e)$ and quarks $(u, d)$.

Within the notation of Ref. [6], the covariant derivative of the fermion field belonging to the representation of $SU(2)$ and having the charge $Y$ with respect to $U(1)$ equals

$$D_\mu = \partial_\mu - igA_\mu^a T_w^a - ig'Y_w B_\mu, \qquad (1)$$

where $A_\mu^a$ and $B_\mu$ are gauge bosons of $SU(2)$ and $U(1)$, respectively.
For the GWS model,

$$D_\mu = \partial_\mu - i\frac{g}{\sqrt{2}}\left(W_\mu^+ T_\mu^+ + W_\mu^- T_\mu^-\right) - i\frac{g}{\cos\theta_w} Z_\mu \left(T_w^3 \sin^2\theta_w Q\right) - ieA_\mu Q. \qquad (2)$$

In expression (2), $W_\mu^\pm, Z_\mu$ are massive vector bosons; $A_\mu$ is the electromagnetic field; $g = \frac{e}{\sin\theta_w}$; $\theta_w$ is the angle of weak mixing; $T_\mu^\pm = T_w^1 \pm iT_w^2, T_w^3$ are weak isospin components; $Q = T_w^3 + Y_w, Y_w$ is the weak hypercharge; $Q \cdot e$ is the electric charge.

In the GWS model, the Lagrangian is written for massless fermions, and masses appear in fermions only upon spontaneous breaking of $SU(2)$ symmetry and introduction of Yukawa interaction between Higgs bosons and fermions.

Considering the further transition to the isotopic Foldy-Wouthuysen representation, let us write the GWS Lagrangian with originally massive fermions without summands responsible for the Yukawa interaction between Higgs bosons and fermions:

$$\mathcal{L} = \bar{\psi}_{v_e}\left(i\not{\partial} - m_{v_e}\right)\psi_{v_e} + \bar{\psi}_e\left(i\not{\partial} - m_e\right)\psi_e +$$
$$+ \bar{\psi}_u\left(i\not{\partial} - m_u\right)\psi_u + \bar{\psi}_d\left(i\not{\partial} - m_d\right)\psi_d + \quad (3)$$
$$+ g\left(W_\mu^+ J_W^{\mu+} + W_\mu^- J_W^{\mu-} + Z_\mu^0 J_Z^\mu\right) + eA_\mu J_{EM}^\mu,$$

where

$$J_W^{\mu+} = \frac{1}{\sqrt{2}}\left(\bar{\psi}_{v_e}\gamma^\mu \frac{1}{2}(1-\gamma^5)\psi_e + \bar{\psi}_u\gamma^\mu \frac{1}{2}(1-\gamma^5)\psi_d\right);$$

$$J_W^{\mu-} = \frac{1}{\sqrt{2}}\left(\bar{\psi}_e\gamma^\mu \frac{1}{2}(1-\gamma^5)\psi_{v_e} + \bar{\psi}_d\gamma^\mu \frac{1}{2}(1-\gamma^5)\psi_u\right);$$

$$J_Z^\mu = \frac{1}{\cos\theta_W}\left[\bar{\psi}_{v_e}\gamma^\mu \frac{1}{2}(1-\gamma^5)\psi_{v_e} + \bar{\psi}_e\gamma^\mu\left(-\frac{1}{2}+\sin^2\theta_W\right)\frac{1}{2}(1-\gamma^5)\psi_e +\right.$$
$$+ \bar{\psi}_e\gamma^\mu\left(\sin^2\theta_W\right)\left(\frac{1}{2}\right)(1+\gamma^5)\psi_e + \bar{\psi}_u\gamma^\mu\left(\frac{1}{2}-\frac{2}{3}\sin^2\theta_W\right)\frac{1}{2}(1-\gamma^5)\psi_u + \quad (4)$$
$$+ \bar{\psi}_u\gamma^\mu\left(-\frac{2}{3}\sin^2\theta_W\right)\frac{1}{2}(1+\gamma^5)\psi_u + \bar{\psi}_d\gamma^\mu\left(-\frac{1}{2}+\frac{1}{3}\sin^2\theta_W\right)\frac{1}{2}(1-\gamma^5)\psi_d +$$
$$\left. + \bar{\psi}_d\gamma^\mu\left(\frac{1}{3}\sin^2\theta_W\right)\frac{1}{2}(1+\gamma^5)\psi_d \right];$$

$$J_{EM}^\mu = \bar{\psi}_e\gamma^\mu(-1)\psi_e + \bar{\psi}_u\gamma^\mu\left(\frac{2}{3}\right)\psi_u + \bar{\psi}_d\gamma^\mu\left(-\frac{1}{3}\right)\psi_d.$$

In expressions (3), (4), at this stage of consideration, we suppose that there is a right and a left neutrino with mass $m_{v_e}$. To ensure consistency with the GWS model after the transition to the IFW representation, the right component of the neutrino field is taken equal to zero, $(\psi_{v_e})_R = 0$.

Lagrangian (3) in the Dirac representation does not possess the $SU(2)$ symmetry owing to the terms of fermion masses that mix up right and left components of fermion fields.

Based upon the formalism of the transition to the isotopic Foldy-Wouthuysen representation [1], we originally derive a Hamiltonian written in terms of the left and right components of fermion fields from Lagrangian (3):

$$\mathcal{H} = \sum_{i=v_e,e,u,d}\left[(\psi_i^\dagger)_L \boldsymbol{\alpha}\mathbf{p}(\psi_i)_L + (\psi_i^\dagger)_R \boldsymbol{\alpha}\mathbf{p}(\psi_i)_R + (\psi_i^\dagger)_L \beta m_i(\psi_i)_R + (\psi_i^\dagger)_R \beta m_i(\psi_i)_L\right] +$$
$$+ g\left(W_\mu^+(J_W^{\mu+})_L + W_\mu^-(J_W^{\mu-})_L + Z_\mu^0(J_Z^\mu)_L + Z_\mu^0(J_Z^\mu)_R\right) + eA_\mu(J_{EM}^\mu)_L + eA_\mu(J_{EM}^\mu)_R, \quad (5)$$

where

$$(J_W^{\mu+})_L = \frac{1}{\sqrt{2}}\left((\psi_{v_e}^\dagger)_L \alpha^\mu(\psi_e)_L + (\psi_u^\dagger)_L \alpha^\mu(\psi_d)_L\right);$$

$$(J_W^{\mu-})_L = \frac{1}{\sqrt{2}}\left((\psi_e^\dagger)_L \alpha^\mu(\psi_{v_e})_L + (\psi_d^\dagger)_L \alpha^\mu(\psi_u)_L\right);$$

$$\left(J_Z^\mu\right)_L = \frac{1}{\cos\theta_w}\left[\left(\psi_{v_e}^\dagger\right)_L \alpha^\mu \left(\psi_{v_e}\right)_L + \left(\psi_e^\dagger\right)_L \alpha^\mu \left(-\frac{1}{2}+\sin^2\theta_w\right)\left(\psi_e\right)_L + \right.$$
$$+\left(\psi_u^\dagger\right)_L \alpha^\mu \left(\frac{1}{2}-\frac{2}{3}\sin^2\theta_w\right)\left(\psi_u\right)_L + \qquad(6)$$
$$\left.+\left(\psi_d^\dagger\right)_L \alpha^\mu \left(-\frac{1}{2}+\frac{1}{3}\sin^2\theta_w\right)\left(\psi_d\right)_L\right];$$

$$\left(J_Z^\mu\right)_R = \frac{1}{\cos\theta_w}\left[\left(\psi_e^\dagger\right)_R \alpha^\mu \sin^2\theta_w \left(\psi_e\right)_R + \left(\psi_u^\dagger\right)_R \alpha^\mu \left(-\frac{2}{3}\sin^2\theta_w\right)\left(\psi_u\right)_R + \right.$$
$$\left.+\left(\psi_d^\dagger\right)_R \alpha^\mu \frac{1}{3}\sin^2\theta_w \left(\psi_d\right)_R\right];$$

$$\left(J_{EM}^\mu\right)_L = \left(\psi_e^\dagger\right)_L \alpha^\mu(-1)\left(\psi_e\right)_L + \left(\psi_u^\dagger\right)_L \alpha^\mu\left(\frac{2}{3}\right)\left(\psi_u\right)_L + \left(\psi_d^\dagger\right)_L \alpha^\mu\left(-\frac{1}{3}\right)\left(\psi_d\right)_L;$$
$$\left(J_{EM}^\mu\right)_R = \left(\psi_e^\dagger\right)_R \alpha^\mu(-1)\left(\psi_e\right)_R + \left(\psi_u^\dagger\right)_R \alpha^\mu\left(\frac{2}{3}\right)\left(\psi_u\right)_R + \left(\psi_d^\dagger\right)_R \alpha^\mu\left(-\frac{1}{3}\right)\left(\psi_d\right)_R.$$

Then, for each fermion field, we introduce eight-component spinors

$$\left(\Phi_i\right)_I = \begin{pmatrix} \left(\psi_i\right)_R \\ \left(\psi_i\right)_L \end{pmatrix}, \quad i = v_e, e, u, d \qquad(7)$$

and a special isotopic space with matrices $\tau_1, \tau_2, \tau_3$ affecting only the four upper and four lower components of the spinors $\left(\Phi_i\right)_I$ (7).

The introduced isotopic space is in no way connected with the weak-isospin space in the Standard Model and, in particular, in the GWS model. Now, Hamiltonian (5) can be written as

$$\mathcal{H} = \sum_{i=v_e,e,u,d}\left\{\left(\Phi_i^+\right)_I\left(\boldsymbol{\alpha}\mathbf{p}+\tau_1\beta m_i\right)\left(\Phi_i\right)_I + e\left(\Phi_i^+\right)_I Q_i \alpha^\mu A_\mu\left(\Phi_i\right)_I + \right.$$
$$\left.+\frac{g}{\cos\theta_w}\left(\Phi_i^+\right)_I\left(T_{Iiw}^3 - \sin^2\theta_w Q_i\right)\alpha^\mu Z_\mu^0\left(\Phi_i\right)_I\right\} - \qquad(8)$$
$$-\frac{g}{\sqrt{2}}\left[\left(\Phi_{v_e}^+\right)_I 2T_{Iv_ew}^3 \alpha^\mu W_\mu^+ 2T_{Iew}^3\left(\Phi_e\right)_I + \left(\Phi_u^+\right)_I 2T_{Iuw}^3 \alpha^\mu W_\mu^+ 2T_{Idw}^3\left(\Phi_d\right)_I\right] +$$
$$-\frac{g}{\sqrt{2}}\left[\left(\Phi_e^+\right)_I 2T_{Iew}^3 \alpha^\mu W_\mu^- 2T_{Iv_ew}^3\left(\Phi_{v_e}\right)_I + \left(\Phi_d^+\right)_I 2T_{Idw}^3 \alpha^\mu W_\mu^- 2T_{Iuw}^3\left(\Phi_u\right)_I\right].$$

In expression (8) $T_{Iiw}^3 = \begin{pmatrix} T_{iR}^3 & 0 \\ 0 & T_{iL}^3 \end{pmatrix}$ is an eight-component weak-isospin matrix with $T_{iR}^3 = 0$ for right particles and $T_{v_e}^3 = \frac{1}{2}, T_{eL}^3 = -\frac{1}{2}; T_{uL}^3 = \frac{1}{2}; T_{dL}^3 = -\frac{1}{2}$ for left particles in accordance with the GWS model. The quantity $Q_i$, which equals $Q_{v_e} = 0, Q_e = -1; Q_u = \frac{2}{3}; Q_d = -\frac{1}{3}.$

is the same for right and left particles. The following equations for fermion fields $\left(\Phi_i\right)_I$ are deduced from Hamiltonian (8):

$$p_0\left(\Phi_{v_e}\right)_I = \left(\boldsymbol{\alpha}\mathbf{p}+\tau_1\beta m_{v_e}+\frac{g}{\cos\theta_w}T_{Iv_ew}^3\alpha^\mu Z_\mu^0\right)\left(\Phi_{v_e}\right)_I - \frac{g}{\sqrt{2}}2T_{Iv_ew}^3\alpha^\mu W_\mu^+ 2T_{Iew}^3\left(\Phi_e\right)_I \quad(9)$$

$$p_0(\Phi_e)_1 = \left(\boldsymbol{\alpha}\mathbf{p} + \tau_1\beta m_e - e\alpha^\mu A_\mu + \frac{g}{\cos\theta_w}\left(T^3_{1ew} + \sin^2\theta_w\right)\alpha^\mu Z^0_\mu\right)(\Phi_e)_1 + $$
$$-\frac{g}{\sqrt{2}}2T^3_{1ew}\alpha^\mu W^-_\mu 2T^3_{1v_e w}(\Phi_{v_e})_1 \tag{10}$$

$$p_0(\Phi_u)_1 = \left(\boldsymbol{\alpha}\mathbf{p} + \tau_1\beta m_u + \frac{2}{3}e\alpha^\mu A_\mu + \frac{g}{\cos\theta_w}\left(T^3_{1uw} - \frac{2}{3}\sin^2\theta_w\right)\alpha^\mu Z^0_\mu\right)(\Phi_u)_1 - $$
$$-\frac{g}{\sqrt{2}}2T^3_{1uw}\alpha^\mu W^+_\mu 2T^3_{1dw}(\Phi_d)_1 \tag{11}$$

$$p_0(\Phi_d)_1 = \left(\boldsymbol{\alpha}\mathbf{p} + \tau_1\beta m_d - \frac{1}{3}e\alpha^\mu A_\mu + \frac{g}{\cos\theta_w}\left(T^3_{1dw} + \frac{1}{3}\sin^2\theta_w\right)\alpha^\mu Z^0_\mu\right)(\Phi_d)_1 + $$
$$-\frac{g}{\sqrt{2}}2T^3_{1dw}\alpha^\mu W^-_\mu 2T^3_{1uw}(\Phi_u)_1 \tag{12}$$

In each of Eqs. (9) – (12), the only summand that mixes up right and left field components is the fermion mass term $(\tau_1\beta m_i)$.

In each Eq. (9) – (12), the interaction between the left fermion field and $W^\pm_\mu$ bosons results in the presence of another left fermion field from the corresponding $SU(2)$ doublet.

For right fermion fields, there is no interaction with $W^\pm_\mu$ bosons in the GWS model.

In accordance with Ref. [1], the fermion fields

$$(\Phi_i)_2 = \tau_1(\Phi_i)_1 = \begin{pmatrix}(\psi_i)_L \\ (\psi_i)_R\end{pmatrix}, \quad i = v_e, e, u, d \tag{13}$$

are also solutions to Eqs. (9) – (12) with a modified weak-isospin matrix

$$T^3_{2iw} = \tau_1 T^3_{1iw}\tau_1 = \begin{pmatrix}T^3_{iL} & 0 \\ 0 & T^3_{iR}\end{pmatrix}. \tag{14}$$

If we introduce 16-component fermion fields

$$\Phi_i = \begin{pmatrix}(\Phi_i)_1 \\ (\Phi_i)_2\end{pmatrix} = \begin{pmatrix}(\psi_i)_R \\ (\psi_i)_L \\ (\psi_i)_L \\ (\psi_i)_R\end{pmatrix}, \quad i = v_e, e, u, d \tag{15}$$

with the replacement of

$$e \to \frac{1}{2}e\left(E_{16\times 16} + \alpha^I_1\right),$$
$$g \to \frac{1}{2}g\left(E_{16\times 16} + \alpha^I_1\right), \tag{16}$$

it will be possible to write equations similar to Eqs. (9) – (12) for the fields $\Phi_i$ (see [1]).

The equations for $(\Phi_i)_1$, $(\Phi_i)_2$, $(\Phi_i)$ are equivalent to each other. However, in the IFW representation, these equations represent different physical patterns [1].

Let us perform the isotopic Foldy-Wouthuysen transformation for the case of the fermion fields $(\Phi_i)_I$ governed by Eqs. (9) – (12).

$$(\Phi_i)_I = (U_i)^+_{IFW} (\Phi_i)_{IIFW}. \tag{17}$$

Each of Eqs. (9) – (12) is then multiplied on the left by the transformation matrix

$$(U_i)_{IFW} = (1 + \delta_{1i} + \delta_{2i} + ...)(U_{0i})_{IFW}. \tag{18}$$

The matrix $U_{IFW}$ is unitary, and expansion in (18) is performed in powers of corresponding coupling constants. The explicit form of the operators $(U_0)_{IFW}$, $\delta_{1i}$, $\delta_{2i}$, ... is defined in Ref. [1].

The IFW transformation gives the following equations:

$$p_0 (\Phi_i)_{IIFW} = (\tau_3 E_i + K_{1i} + K_{2i} + ...)(\Phi_i)_{IIFW} + \\ + (K'_{1i \to j} + K'_{2i \to j} + ...)(\Phi_j)_{IIFW};\ i,j = v_e, e\ \text{или}\ u, d;\ i \neq j. \tag{19}$$

Eqs. (19) contain equations for the lepton and quark $SU(2)$ left doublets and right singlets, respectively.

In Eqs. (19), in the expressions for $K_{1i}, K_{2i}$ ... describing the electromagnetic interaction and weak interactions with exchange of $Z^0_\mu$ bosons, the $q\alpha_\mu B^\mu$ interaction, which is used in Ref. [1], should be replaced as follows:

$$q\alpha_\mu B^\mu \to eQ_i \alpha^\mu A_\mu + \frac{g}{\cos\theta_w}(T^3_{1iw} - \sin^2\theta_w\ Q_i)\alpha^\mu Z^0_\mu. \tag{20}$$

In Eqs. (19), the summands $K'_{1i \to j}, K'_{2i \to j},...$ are responsible for the weak interaction involving charged $W^\pm$ bosons. Let us consider the structure of expressions for $K'_{ni \to j}$ by the example of the transformation of Eq. (9). After the IFW transformation, the last summand in Eq. (9) takes the following form:

$$(U_{v_e})_{IFW}\left(-\frac{g}{\sqrt{2}} 2T^3_{1v_e w}\alpha^\mu W^+_\mu 2T^3_{1ew}\right)(U^+_e)_{IFW}(\Phi_e)_{IIFW} = \\
= (1 + \delta_{1v_e} + \delta_{2v_e} + ...)(U^+_{0e})_{IFW}\left(-\frac{g}{\sqrt{2}} 2T^3_{1v_e w}\alpha^\mu W^+_\mu 2T^3_{1ew}\right)(U^+_{0e})_{IFW}(1 + \delta_{1v_e} + \delta_{2v_e} + ...)(\Phi_e)_{IIFW} = \\
= \left[(U_{0v_e})_{IFW}\left(-\frac{g}{\sqrt{2}} 2T^3_{1v_e w}\alpha^\mu W^+_\mu 2T^3_{1ew}\right)(U^+_{0e})_{IFW} + \delta_{1v_e}(U_{0v_e})_{IFW}\left(-\frac{g}{\sqrt{2}} 2T^3_{1v_e w}\alpha^\mu W^+_\mu 2T^3_{1ew}\right)(U^+_{0e})_{IFW} - \right. \\
\left. -(U_{0v_e})_{IFW}\left(-\frac{g}{\sqrt{2}} 2T^3_{1v_e w}\alpha^\mu W^+_\mu 2T^3_{1ew}\right)(U^+_{0e})_{IFW} \delta_{1e} + ...\right](\Phi_e)_{IIFW} = (K'_{1v_e \to e} + K'_{2v_e \to e} + ...)(\Phi_e)_{IIFW}. \tag{21}$$

In expression (21), the operators $\delta_{1v_e}$ and $\delta_{1e}$ are defined within the formalism of Ref. [1] with replacement (20).

In Eqs. (19), the summands with $\tau_3 E_i, K_{ni}$, by definition, are even with respect to the mixing-up of upper (right) and lower (left) components $(\Phi_i)_{IIFW}$. The summands with $K'_{nv_e \to e}$ are formally odd, but they actually can be treated as even because only left fermions participate in the interaction with charged bosons $W_\mu^\pm$, which is reflected in the definition of $T^3_{1ew}$ (see (8)).

For the fermion fields $(\Phi_i)_{2FW}$, one can write equations similar to (19) with replacement of $T^3_{1iw} \to T^3_{2iw}$ (see (14)). Finally, one can derive Foldy-Wouthuysen equations in the IFW representation for Dirac fields $\Phi_i$ (15) with replacement (16) in expressions (20), (21).

Let us write these equations together with Eqs. (19) and their Hamiltonian densities.

$$p_0(\Phi_i)_{1IFW} = (\tau_3 E_i + K_{1i} + K_{2i} + ...)(\Phi_i)_{1IFW} + \\ + (K'_{1i \to j} + K'_{2i \to j} + ...)(\Phi_j)_{1IFW};$$
(22)

$$\mathcal{H}'_{IFW} = \sum_i (\Phi_i)^+_{1IFW}(\tau_3 E_i + K_{1i} + K_{2i} + ...)(\Phi_i)_{1IFW} + \\ + \sum_{i,j}(\Phi_i)^+_{1IFW}(K'_{1i \to j} + K'_{2i \to j} + ...)(\Phi_j)_{1IFW};$$
(23)

$$p_0(\Phi_i)_{2IFW} = (\tau_3 E_i + K_{1i} + K_{2i} + ...)(\Phi_i)_{2IFW} + \\ + (K'_{1i \to j} + K'_{2i \to j} + ...)(\Phi_j)_{2IFW};$$
(24)

$$\mathcal{H}''_{IFW} = \sum_i (\Phi_i)^+_{2IFW}(\tau_3 E_i + K_{1i} + K_{2i} + ...)(\Phi_i)_{2IFW} + \\ + \sum_{i,j}(\Phi_i)^+_{2IFW}(K'_{1i \to j} + K'_{2i \to j} + ...)(\Phi_j)_{2IFW};$$
(25)

$$P_0(\Phi_i)_{IFW} = (\tau_3 E_i + K^\Phi_{1i} + K^\Phi_{2i} + ...)(\Phi_i)_{IFW} + \\ + (K'^\Phi_{1i \to j} + K'^\Phi_{2i \to j} + ...)(\Phi_j)_{IFW};$$
(26)

$$\mathcal{H}'''_{IFW} = \sum_i (\Phi_i)^+_{IFW}(\tau_3 E_i + K^\Phi_{1i} + K^\Phi_{2i} + ...)(\Phi_i)_{IFW} + \\ + \sum_{i,j}(\Phi_i)^+_{IFW}(K'^\Phi_{1i \to j} + K'^\Phi_{2i \to j} + ...)(\Phi_j)_{IFW}.$$
(27)

In expressions (22) – (27), $i, j = v_e, e$ and $u, d; i \neq j$; in expressions (22), (23) $T^3_{1iw} = \begin{pmatrix} T^3_{iR} & 0 \\ 0 & T^3_{iL} \end{pmatrix}$; in expressions (24), (25) $T^3_{2iw} = \begin{pmatrix} T^3_{iL} & 0 \\ 0 & T^3_{iR} \end{pmatrix}$; in expressions (26), (27), we use the matrix $T^3_{1iw}$ for the eight-component fields $(\Phi_i)_{1IFW}$ and the matrix $T^3_{2iw}$ for the fields $(\Phi_i)_{2IFW}$.

Eqs. (26) and Hamiltonian (27) are basically in line with the description of elementary particles and their interactions in the GWS model. To ensure complete compliance, one should exclude the states with the right neutrino from (26) - (27). At this stage, this can be done easily by taking $(\psi_{v_e})_R = 0$ in the basis functions $(\Phi_{v_e})_{1IFW}, (\Phi_{v_e})_{2IFW}, (\Phi_{v_e})_{IFW}$.

Eqs. (22) and Hamiltonian (23) describe the motion and interactions of right fermions and left antifermions with lack of interactions between real particles and antiparticles.

Eqs. (24) and Hamiltonian (25) on the contrary describe the motion and interactions of left fermions and right antifermions, also with lack of interactions between real particles and antiparticles.

The three physical patterns described above are discussed in more detail in Ref. [1].

Equations and Hamiltonians (22) – (27) in the IFW representation are actually written in the chirally symmetric form, whether fermions have mass or not. Expressions (22) – (27) are invariant with respect to the transformation of chiral symmetry

$$
\begin{aligned}
(\Phi_i)_{1IFW} &\to e^{i\alpha\gamma^5}(\Phi_i)_{1IFW}; \\
(\Phi_i)_{2IFW} &\to e^{i\alpha\gamma^5}(\Phi_i)_{2IFW}; \\
(\Phi_i)_{IFW} &\to e^{i\alpha\gamma^5}(\Phi_i)_{IFW}.
\end{aligned}
\quad (28)
$$

Note that the operator $\gamma^5$ in the IFW representation is not connected with helicity of fermions. This connection exists for the operator $(\gamma^5)_{IFW} = U_{IFW}\gamma^5 U_{IFW}^+$ (see [1]).

Taking the foregoing into account, equations and Hamiltonians (22) - (27) are $SU(2)$-invariant, whether fermions have mass or not. Consequently, the $SU(2)$-invariant formulation of the GWS model in the isotopic Foldy-Wouthuysen representation exists for both massless and massive fermions. This conclusion is not changed neither by accounting for two other generations of particles in the Standard Model $(v_\mu,\mu,c,s)$ and $(v_\tau,\tau,t,b)$ in the IFW representation, nor by accounting for $SU(3)$ - symmetric quantum chromodynamics.

Thus, we demonstrated that it is not necessary for ensuring of $SU(2)$-invariance of theory to introduce of Yukawa interaction between Higgs bosons and fermions in the isotopic Foldy-Wouthuysen representation.

## 2. SPONTANEOUS BREAKING OF PARITY IN THE IFW REPRESENTATION AND PROBLEMS OF "DARK MATTER"

As already discussed [1], three vacuum states correspond to chirally symmetric equations and Hamiltonians (22) – (27). The ground state of Hamiltonian (27) is a Dirac sea of right and left negative-energy fermions. The ground state of Hamiltonians (23), (25) is a sea of left and a sea of right negative-energy fermions, respectively.

It is clear that there are prerequisites for spontaneous breaking of parity and transition from the vacuum of Hamiltonian (27) to the vacuum of Hamiltonian (23) or Hamiltonian (25).

Without proposing any specific mechanism breaking of P - symmetry, the author points out that in the case of such breaking, the description of Nature by Eqs. and Hamiltonians (22) – (25) is significantly poorer than its description by expressions (26), (27). Expressions (22) – (25) describing either only right fermions and left antifermions, or only left fermions and right antifermions exclude strong and electromagnetic interactions because of their P-, C-invariance. In expressions (22) – (23) describing the motion and interactions of right fermions and left antifermions, there remains the possibility of weak interactions through the neutral current of right particles $\left(J_Z^\mu\right)_R$ (see 6). In expressions (24), (25) describing the motion and interactions of left fermions and right antifermions, weak interactions occur through charged currents of left particles $\left(J_W^{\mu+}\right)_L$, $\left(J_W^{\mu-}\right)_L$ and through the neutral current of left particles $\left(J_Z^\mu\right)_L$. Note that in both cases Eqs. and Hamiltonians (22) – (25) do not include processes with simultaneous involvement of real particles and antiparticles due to the spinor structure of the basis functions [1]. These processes involve a direct and reverse $\beta^-$ decay etc.

Thus, Eqs. and Hamiltonians (22) – (25) describe the motion of left and right particles of the Standard Model having no strong and electromagnetic interactions and participating only in weak interactions by strongly restricted channels described above.

The world corresponding to the physical patterns described by expressions (22), (23), or by expressions (24), (25), should possess the following properties:

- it should not emit and absorb light;
- it should look electrically neutral;
- motion of particles should be non-relativistic;
- it should weakly interact with the outside world;
- quarks should move without confinement.

In literature, the listed properties (except for the latter) are basic properties ascribed to cold "dark matter" (see, e.g., [7]). Based on this, one can suppose that "dark matter" came into existence at a certain time and in a certain part of the Universe as a result of spontaneous breaking of parity. This resulted in the transition from the world described by Eqs. and Hamiltonian (26), (27) to the world described by expressions (22), (23) or expressions (24), (25). In this case, "dark matter" consists of either right fermions and left antifermions, or left fermions and tight antifermions with the particle composition corresponding to the Standard Model. Within this approach, "dark matter" contains free quarks moving without confinement.

"Dark matter" particles interact with each other and the world of "light matter" through restricted channels of weak interaction without simultaneous participation of real particles and antiparticles in the processes.

## 2. CONCLUSIONS

The paper formulates the Standard Model with massive fermions in the isotopic Foldy-Wouthuysen representation. Equations of fermion fields and their Hamiltonians in the IFW representation are invariant with respect to the transformations of chiral symmetry. $SU(2) \times U(1)$ invariance of the theory in the IFW representation is independent of whether fermions possess mass or not. It follows from this that it is not necessary to introduce interactions between Higgs bosons and fermions. In this case, fermion masses are introduced from outside. Higgs bosons are responsible for the gauge invariance of the theory's boson sector and interact only with gauge vector bosons $W_\mu^\pm$, $Z_\mu^0$, gluons and photons. Within this approach, there are no processes of Higgs boson decay to fermions $(H \to f \bar{f})$, no quarkonium states $\psi, \gamma, \theta$ including Higgs bosons, no interactions between Higgs bosons and gluons $(ggH)$ and photons $(\gamma\gamma H)$ via fermion loops, etc.

Because of the unitarity of the Foldy-Wouthuysen representation, all the other theoretical results of the Standard Model should be the same as the results obtained earlier in the Dirac representation.

The second part of this study explores the possible relation between spontaneous breaking of P - symmetry in the IFW representation and the problems and composition of "dark matter" particles.

If "dark matter" after spontaneous breaking of parity is described by Eqs. and Hamiltonians (22) – (25), it is a set of either left particles and right antiparticles, or right particles and left antiparticles of the Standard Model without interactions of real particles with antiparticles. These sets of particles have no strong and electromagnetic interactions and participate in restricted channels of weak interactions without processes with simultaneous presence of real particles and antiparticles. Within this approach, "dark matter" contains free quarks moving without confinement.

The next step in validating the set of "dark matter" particles proposed in the study is development of mechanism of spontaneous breaking of parity in the isotopic Foldy-Wouthuysen representation and to compare the consequences breaking of P - symmetry with existing experimental data.